\documentclass[reprint,prd,nofootinbib]{revtex4-1}

\usepackage{bm}
\usepackage{graphicx} %
\usepackage{amsmath} %
\usepackage{amssymb} %
\usepackage{url}
\usepackage{subfigure} %
\usepackage{float} %
\usepackage{upgreek} %
\usepackage{multirow} %
\usepackage{color} %
\usepackage{lineno} %
\usepackage{hyperref} %
\usepackage{booktabs}
\usepackage{placeins}
\usepackage[usenames,dvipsnames]{xcolor}

\allowdisplaybreaks[4]

\makeatletter
\DeclareFontFamily{OMX}{MnSymbolE}{}
\DeclareSymbolFont{MnLargeSymbols}{OMX}{MnSymbolE}{m}{n}
\SetSymbolFont{MnLargeSymbols}{bold}{OMX}{MnSymbolE}{b}{n}
\DeclareFontShape{OMX}{MnSymbolE}{m}{n}{
    <-6>  MnSymbolE5
   <6-7>  MnSymbolE6
   <7-8>  MnSymbolE7
   <8-9>  MnSymbolE8
   <9-10> MnSymbolE9
  <10-12> MnSymbolE10
  <12->   MnSymbolE12
}{}
\DeclareFontShape{OMX}{MnSymbolE}{b}{n}{
    <-6>  MnSymbolE-Bold5
   <6-7>  MnSymbolE-Bold6
   <7-8>  MnSymbolE-Bold7
   <8-9>  MnSymbolE-Bold8
   <9-10> MnSymbolE-Bold9
  <10-12> MnSymbolE-Bold10
  <12->   MnSymbolE-Bold12
}{}

\let\llangle\@undefined
\let\rrangle\@undefined
\DeclareMathDelimiter{\llangle}{\mathopen}%
                     {MnLargeSymbols}{'164}{MnLargeSymbols}{'164}
\DeclareMathDelimiter{\rrangle}{\mathclose}%
                     {MnLargeSymbols}{'171}{MnLargeSymbols}{'171}
\makeatother

\hypersetup{colorlinks=true, urlcolor=black, linkcolor=blue, citecolor=red} %

\begin{document}

\title{CMB spectroscopy at third-order in cosmological perturbations}

\author {Atsuhisa Ota${}^1$}
\email{a.ota@uu.nl}
\author {Nicola Bartolo${}^{2,3,4}$}
\affiliation{${}^1$Institute for Theoretical Physics and Center for Extreme Matter and Emergent Phenomena,
Utrecht University,\\ Princetonplein 5, 3584 CC Utrecht, The Netherlands}
\affiliation{${}^2$Dipartimento di Fisica e Astronomia ``G. Galilei'', Universit\`a degli Studi di Padova, via Marzolo 8, I-35131, Padova, Italy}
\affiliation{${}^3$INFN, Sezione di Padova, via Marzolo 8, I-35131, Padova, Italy}
\affiliation{${}^4$INAF-Osservatorio Astronomico di Padova, Vicolo dell'Osservatorio 5, I-35122 Padova, Italy}

\date{\today}

\begin{abstract}

Early energy injection to the Cosmic Microwave Background~(CMB) from dissipation of acoustic waves generates deviations from the blackbody spectrum not only at second-order but also at third-order in cosmological perturbations.
We compute this new spectral distortion $\mathcal \kappa$ based on third-order cosmological perturbation theory and show that $\kappa$ arises as a result of mode coupling between spectral distortions and temperature perturbations.
The ensemble average of $\kappa$ can be directly sourced by (integrated) primordial non-Gaussianity.
In particular, we roughly estimate the signal as $\kappa=f^{\rm loc.}_{\rm NL}\times \mathcal O(10^{-18})$ for local type scale-independent non-Gaussianity.
The signal is incredibly tiny; however, we argue that it carries a specific frequency dependence different from other types of CMB spectral distortions. Also, it should be noticed that $\kappa$ is sensitive to extremely squeezed shapes of primordial bispectra that cannot be constrained by the CMB anisotropies. Finally, we comment on other possible applications of our results.


\end{abstract}

\maketitle


\section{Introduction}
Distortions to the blackbody spectrum of the Cosmic Microwave Background~(CMB) from dissipation of acoustic waves have been intensely investigated to study primordial density perturbations which provide us with rich information on cosmic inflation~\cite{inflation,Sunyaev:1970eu,Hu:1994bz,Chluba:2011hw,Chluba:2012we,Chluba:2012gq}.
The effect is known to be second-order in the cosmological perturbations; therefore, the ensemble averages of the distortions directly arise from the primordial power spectrum, and its anisotropy can be related to the primordial bispectrum~(i.e., primordial non-Gaussianity)~\cite{Pajer:2012vz,Ganc:2012ae,Pajer:2013oca,Biagetti:2013sr,Ota:2014iva,Emami:2015xqa,Shiraishi:2015lma,Bartolo:2015fqz,Bartolo:2015fqz,Shiraishi:2016hjd,Ravenni:2017lgw,Chluba:2016aln,Khatri:2015tla}. The CMB spectral distortions are usually classified into two types:   
$\mu$ and $y$, the chemical potential and the Compton $y$ parameter, respectively~(see Refs.~\cite{Khatri:2012tw,Chluba:2013pya,Khatri:2013dha} for other types of spectral distortions). 
The monopole of the $\mu$- and the $y$-distortions from damping of short wavelength acoustic waves can be estimated as $10^{-8}$ and $10^{-9}$ for almost scale-invariant Gaussian adiabatic perturbations, and they are one of the targets of next generation of space missions~\cite{Kogut:2011xw,Andre:2013afa}.
Thus, the CMB spectral distortions are known as a powerful tool for observations of the primordial density perturbations on small scales.
In this paper, we point out another spectral distortion from dissipation of acoustic waves at third-order in the cosmological perturbations, and that its ensemble average can be directly sourced by primordial non-Gaussianity.  Since it is third-order in the cosmological perturbations, the signal can be thought of as tiny.
Still, in principle, we can distinguish it from the other types of spectral distortions such as $\mu$ and $y$ because of its peculiar frequency dependence.
In this paper, we compute such a third-order spectral distortion in the early Universe for the first time.
\\


\section{Formalism}
\label{sec:formalism}


\subsection{Set up}
The CMB radiation initially follows the local blackbody spectrum due to frequent interactions.
However, deviations from the local blackbody spectrum are possible, e.g.,  for the redshift $z \lesssim 5 \times 10^4$.
During this epoch, the Compton scattering is too weak against Hubble expansion to establish local kinetic equilibrium states so that $y$-distortions are generated~\cite{ddz}.
One linearizes the photon Boltzmann equations to find the evolution of the temperature perturbations.
The $y$-distortion is a deviation from the local blackbody spectrum that appears at the next-to-leading order in the cosmological perturbations.
More generally, we introduce the following ansatz for the photon Boltzmann equation up to third-order~\cite{Ota:2016esq}:
\begin{align}
f(\eta,\mathbf x, p\mathbf n)=\frac{1}{e^{\frac{p}{T_{\rm rf}}e^{-\Theta}}-1}+y\mathcal Y(p)+\kappa \mathcal K(p),\label{dist:decompose}
\end{align}
where $(\eta,\mathbf x)$ are \textit{comoving} spacetime coordinates, $p$ is the magnitude of the photon \textit{comoving} momentum, $\mathbf n$ is photon's direction, $T_{\rm rf}=2.725$K is the temperature of the comoving blackbody.
The temperature perturbation $\Theta$, the $y$-distortion $y$, the new third-order distortion $\kappa$ are functions of $(\eta, \mathbf x,\mathbf n)$: they are $p$ independent.
These parameters can be expanded perturbatively as 
\begin{align}
	\Theta &= \Theta^{(1)} + \Theta^{(2)}+ \Theta^{(3)}+\cdots,\\
		y&=y^{(2)}+ y^{(3)}+\cdots,\\
	\kappa &= \kappa^{(3)}+\cdots,
\end{align}
with superscripts being the order of the cosmological perturbations.
We have also defined the momentum basis 
\begin{align}
	f^{(0)}(p)&\equiv\frac{1}{e^{{\frac{p}{T_{\rm rf}}}}-1},\\
	\mathcal G(p) &\equiv \left(-p~\frac{\partial}{\partial p}\right)f^{(0)},\\
	\mathcal Y(p) &\equiv \left(-p~\frac{\partial}{\partial p}\right)^{2}f^{(0)}-3\mathcal G,\\
	\mathcal K(p) &\equiv \left(-p~\frac{\partial}{\partial p}\right)^{3}f^{(0)}-3\mathcal Y -9\mathcal G.	
\end{align}
Then, all $p$ dependences in Eq.~(\ref{dist:decompose}) can be factorized by these functions. 
This implies that we can in principle distinguish $\kappa$ from $y$ thanks to the difference between $\mathcal K$ and $\mathcal Y$, which are both defined not to change the number of photons~(see Ref.~{\cite{Ota:2016esq} for the details of these functions.).
We have omitted the chemical potential, that is, the $\mu$-distortion $\mu$ because we only consider the late epoch out of kinetic equilibrium for simplicity.
The primary goal of this paper is to derive the evolution equation of this $\kappa$. 

\subsection{Harmonic expansions and primordial random fields}
We introduce a harmonic coefficient of $A(\eta,\mathbf x,\mathbf n)$ as 
\begin{align}
	A_{lm}(\eta,\mathbf x)\equiv \int d\mathbf n Y^{*}_{lm}(\mathbf n)A(\eta,\mathbf x,\mathbf n).
\end{align}
The Fourier integral 
\begin{align}
A(\eta,\mathbf k,\mathbf n)\equiv\int d^{3}xe^{-i\mathbf k\cdot\mathbf x}A(\eta,\mathbf x,\mathbf n),
\end{align}
is linear in the primordial curvature perturbation on the uniform density slice $\zeta_{\mathbf k}$.
We expand it by using the Legendre polynomials as
\begin{align}
    \begin{split}
        A(\eta,\mathbf k,\mathbf n)&=\sum_{l}(-i)^{l}(2l+1)P_l(\mathbf n\cdot \hat k)A_{l}(\eta,k)\zeta_{\mathbf k} \\
&=(4\pi)\sum_{lm}(-i)^{l}Y_{lm}(\mathbf n)Y^{*}_{lm}(\hat k)A_{l}(\eta,k)\zeta_{\mathbf k},\label{Fourier:int}
    \end{split}
\end{align}
where we call $A_{l}(\eta,k)$ ``transfer function'' of $A$.
Note that, in this paper, $A_{lm}$ is always defined in real space. Similarly, $A_l$ is given in Fourier space.
We write the primordial power spectrum and bispectrum of $\zeta$ calculated in inflationary models as~(see, e.g.,~\cite{Acquaviva:2002ud,Maldacena:2002vr,Bartolo:2004if,Chen:2010xka}).
\begin{align}
	\langle \zeta_{\mathbf k_{1}}\zeta_{\mathbf k_{2}}\rangle &= (2\pi)^{3}\delta^{(3)}(\mathbf k_{1}+\mathbf k_{2})P_{\zeta}(k_{1}),\\
	\langle \zeta_{\mathbf k_{1}}\zeta_{\mathbf k_{2}}\zeta_{\mathbf k_{3}}\rangle &= (2\pi)^{3}\delta^{(3)}(\mathbf k_{1}+\mathbf k_{2}+\mathbf k_{3})B_{\zeta}(k_{1},k_{2},k_{3}).
\end{align}

\subsection{Liouville terms}
Thanks to the parametrization~(\ref{dist:decompose}), the Boltzmann equation for the photon distribution function translates into the equations for the coefficients of $\mathcal G$, $\mathcal Y$ and $\mathcal K$~\cite{Ota:2016esq}. 
Expanding Eq.~(\ref{dist:decompose}) up to third-order in cosmological perturbations, one finds 
\begin{align}
f=&f^{(0)}+[\Theta +\cdots]\mathcal G \notag \\
&+\left[y+\frac{1}{2}\Theta^{2}+\cdots\right]\mathcal Y
+\left[\frac{1}{3!}\Theta^{3}+\kappa\right]\mathcal K,
\end{align}
where the dots imply the next-to-leading order corrections to each part here and hereafter.
We take a derivative of both sides w.r.t. the conformal time to obtain
\begin{align}
&\frac{df}{d\eta}=\left[\frac{d\Theta}{d\eta}-\frac{d \ln p}{d\eta} +\cdots \right]\mathcal G
\notag \\
&+\left[\frac{dy}{d\eta}+\Theta \left(\frac{d\Theta}{d\eta} -\frac{d \ln p}{d\eta}\right)+ \cdots\right]\mathcal Y \notag \\
&+\left[\frac{d\kappa}{d\eta}- y \frac{d\ln p }{d\eta}+ \frac{1}{2}\Theta^{2}\left(\frac{d\Theta}{d\eta} -\frac{d \ln p}{d\eta}\right)
+\cdots \right]\mathcal K,\label{dfdeta2}
\end{align}
where we used 
\begin{align}
\frac{d\mathcal Y}{d\eta}=\frac{dp}{d\eta} \cdot \frac{d\mathcal Y}{dp}=-\frac{d\ln p}{d\eta} \cdot \mathcal K,	
\end{align}
 and one can use similar techniques for $\mathcal G$ and $f^{(0)}$.
Note that $d\ln p/d\eta$ starts with linear perturbations since $p$ is the comoving momentum; therefore, terms with a time derivative of $\mathcal K$ become fourth-order.
The gravitational effects are included in $d\ln p/d\eta$, which does not have any explicit $p$ dependence even at nonlinear order~(see, e.g., Ref.~\cite{Dodelson:2003ft} for the linear case). 
Thus, the $p$ dependence of the Liouville term can be reduced to the linear combination of $\mathcal G$, $\mathcal Y$, and $\mathcal K$.

\subsection{Collision terms for the Compton scattering}
Next, let us consider the right hand side~(RHS) of the Boltzmann equation.
For $z \lesssim 5 \times 10^4$, the collision terms for the Compton scattering can be expanded into the following form up to third-order in the cosmological perturbations~\cite{Ota:2016esq}:
\begin{align}
\mathcal C_{\rm T}[f]=\mathcal A \mathcal G + \mathcal B \mathcal Y+\mathcal D \mathcal K,\label{col:1}
\end{align}
where $\mathcal A=\mathcal A^{(1)}+\cdots$, $\mathcal B=\mathcal B^{(2)}+\cdots$ and $\mathcal D=\mathcal D^{(3)}+\cdots$ are $p$ independent.
We may drop the other linear order corrections with $(1+z)p/m_{\rm e}$, $(1+z)T_{\rm rf}/m_{\rm e}$ and $T_{\rm e}/m_{\rm e}$, where $z$, $T_{\rm e}$ and $m_{\rm e}$ are the redshift, the physical electron temperature and electron mass respectively.
This is because the ensemble average of the linear perturbations are zero so that they do not affect our final expression~\cite{Haga:2018pdl}.
Combining Eqs.~(\ref{dfdeta2}) and (\ref{col:1}), we obtain the following Boltzmann equations for $\Theta$, $y$ and $\kappa$:
\begin{align}
\frac{d\Theta}{d\eta}-\frac{d \ln p}{d\eta} +\cdots &=\mathcal A,\label{temp:bol}\\
\frac{dy}{d\eta}+\Theta \left(\frac{d\Theta}{d\eta} -\frac{d \ln p}{d\eta}\right)+ \cdots&=\mathcal B,\label{y:bol}\\
\frac{d\kappa}{d\eta}- y \frac{d\ln p }{d\eta}+ \frac{1}{2}\Theta^{2}\left(\frac{d\Theta}{d\eta} -\frac{d \ln p}{d\eta}\right)&=\mathcal D\label{k:bol}.
\end{align}

\section{Solving the Boltzmann equations}
\label{sec:Boltzmann}


\subsection{$y$-distortion from acoustic damping}
Before focusing on the third-order distortion, let us derive the evolution equation for the second-order $y$ based on cosmological perturbation theory. 
This can be a useful preliminary computation that provides a term of comparison to the physics giving rise to $\kappa$.
Eqs.~(\ref{temp:bol}) and (\ref{y:bol}) yield
\begin{align}
\frac{dy}{d\eta} =
-\Theta\mathcal A + \mathcal B +\cdots.\label{yevolv}
\end{align}
The leading order terms of $\mathcal A$ are~\cite{Dodelson:2003ft}
\begin{align}
&
-\dot\tau^{-1}\mathcal A=\frac{\Theta_{00}}{\sqrt{4\pi}}-\Theta +V + \frac{1}{10}\sum^{2}_{m=-2}Y_{2m}\Theta_{2m}
,\label{coltemplin}
\end{align}
where $V=\mathbf n \cdot \mathbf v$ with $\mathbf v$ being the velocity of the baryon fluid.
$\tau$ is the optical depth and its dot implies a derivative w.r.t. the conformal time~($\dot\tau<0$).
Those of $\mathcal B$ are~\cite{Chluba:2012gq}
\begin{align}
&-\dot\tau^{-1}\mathcal B
=\frac{y_{00}}{\sqrt{4\pi}} -y
+\frac{1}{10}\sum_{m=-2}^{2}Y_{2m}y_{2m}
\notag \\
&
+\frac{[\Theta^{2}]_{00}}{2\cdot{\sqrt{4\pi}}} -\frac12\Theta^{2}
+\frac{1}{20}\sum_{m=-2}^{2}Y_{2m}[\Theta^{2}]_{2m}
\notag \\
&+\frac{\Theta_{00}}{\sqrt{4\pi}}V-\frac{[V\Theta]_{00}}{\sqrt{4\pi}}+\frac{1}{2}V^2 
+\frac{[V^2]_{00}}{2\cdot \sqrt{4\pi}}\notag \\
&+\frac{1}{10}\sum_{m=-2}^{2}Y_{2m}\bigg[V\Theta_{2m} -[V\Theta]_{2m}+\frac12[V^2]_{2m}\bigg].\label{B2}
\end{align}
Then we obtain the following evolution equation for $y$ up to second-order:
\begin{align}
&-\dot\tau^{-1}\frac{d y}{d\eta}=
\frac{y_{00}}{\sqrt{4\pi}} -y 
+\frac{1}{10}\sum_{m=-2}^{2}Y_{2m} y_{2m}
\notag \\
&-\frac{\Theta_{00}}{\sqrt{4\pi}}\Theta +\Theta^{2} -V\Theta - \frac{1}{10}\Theta\sum^{2}_{m=-2}Y_{2m}\Theta_{2m}  \notag \\
&
+\frac{[\Theta^{2}]_{00}}{2\cdot \sqrt{4\pi}} -\frac12\Theta^{2}
+\frac{1}{20}\sum_{m=-2}^{2}Y_{2m}[\Theta^{2}]_{2m}
\notag \\
&
+\frac{\Theta_{00}}{\sqrt{4\pi}}V -\frac{[V\Theta]_{00}}{\sqrt{4\pi}}+\frac{1}{2}V ^2 
+\frac{[V^2]_{00}}{2\cdot \sqrt{4\pi}}\notag \\
&+\frac{1}{10}\sum_{m=-2}^{2}Y_{2m} \bigg[V \Theta_{2m} -[V\Theta]_{2m}+[V^2]_{2m}\bigg].
\label{yinhomoeq}
\end{align}
The isotropic part of the equation has a simple form:
\begin{align}
&-\dot\tau^{-1}\frac{dy_{00}}{d\eta}
=-\frac{\Theta_{00}^{2}}{\sqrt{4\pi}}+\left[\Theta^{2}\right]_{00}
\notag \\
& -2[V\Theta]_{00}
+[V^{2}]_{00}
+ \frac{1}{10\cdot \sqrt{4\pi}}\sum^{2}_{m=-2}|\Theta_{2m}|^{2}.
\label{realspydif}
\end{align}
Practically, we express the above formula by using the transfer functions in Fourier space calculated by Boltzmann codes.
The theoretical prediction is given by taking the ensemble average using Eq.~(\ref{Fourier:int})~\cite{Chluba:2012gq}:
\begin{align}
\frac{d\llangle y\rrangle}{d\eta} =& 
-\dot\tau
\int \frac{dk}{k}\frac{k^{3} P_{\zeta}(k)}{2\pi^{3}}\left[\frac{9}{2}\Theta^{2}_{2}+ 3\Theta_{1g}^{2}\right],\label{yevol}
\end{align}
where $\Theta_{1g}\equiv \Theta_{1}- V_{1}$ is the gauge invariant relative velocity between photons and baryons, and we drop $l\geq 3$ since the higher order multipoles are less significant due to the exponential damping of higher multipoles during free streaming~\cite{Chluba:2012gq}.
Note that $\llangle \cdots \rrangle$ implies that we take both the ensemble average and the sky average of $\mathbf n$.
Thus, $y$ is related to the primordial power spectrum in a framework of second-order Boltzmann equations.
It is generated from shear viscosity $\Theta_{2}$ and heat conduction $\Theta_{1g}$, which are both gauge invariant at linear order.

\subsection{$\kappa$-distortion from acoustic damping}
Similar steps are possible at third-order, and we naively expect the third-order distortion is directly related to the primordial bispectrum in analogy with Eq.~(\ref{yevol}).
From Eqs.~(\ref{dfdeta2}), (\ref{temp:bol}), (\ref{y:bol}) and (\ref{k:bol}) we obtain
\begin{align}
\frac{d}{d\eta}\left(\kappa-\Theta y \right) =&\frac{1}{2}\Theta^{2}\mathcal A - y \mathcal A  -\Theta\mathcal B  +\mathcal D.\label{kdis:evo}
\end{align}
In contrast to Eq.~(\ref{yevolv}), we find the total derivative $d(\Theta y)/d\eta$.
Since $y=0$ at the initial time, this term turns into a product of $\Theta$ and $y$ at present.
In other words, this part is automatically fixed by $\Theta$ and $y$.
Therefore, it can be thought of as an offset of $\kappa$, and the contribution of physical processes in the early universe is $\bar\kappa = \kappa - \Theta y$.
$\bar \kappa$ also enables us to pin down a gauge-invariant part of $\kappa$ as we will show below~(see also Refs.~\cite{Naruko:2013aaa,Pitrou:2007jy,Bartolo:2006cu,Bartolo:2006fj} for gauge-invariance of spectral distortions).

Let us evaluate the isotropic component of the ensemble average of $\bar \kappa$.
Here we assume the separable form bispectrum for simplicity:
\begin{align}
\langle \zeta_{\mathbf k_{1}}\zeta_{\mathbf k_{2}}\zeta_{\mathbf k_{3}}\rangle
=&\int d^{3}x \sum_{j}\prod_{i=1}^{3}e^{i\mathbf k_{i}\cdot\mathbf x} f^{(ij)}(k_{i}),\label{separable}
\end{align}
which includes, e.g., the ``local'' and ``equilateral''  shapes. 
Hereafter, we frequently take the angle averages and then the ensemble averages of triple products of perturbations calculated in the following way:
\begin{align}
&\int \frac{d \mathbf n}{4\pi}  \left \langle \prod_{i=1}^{3}A_{i}(\eta,\mathbf x,\mathbf n)\right\rangle\notag \\
&=(4\pi)^{2}\int dr r^{2}\sum_{j}\prod_{i=1}^{3}\bigg[\int \frac{dk_{i} k_{i}^{2}}{2\pi^{2}}\sum_{l_{i}m_{i}} A_{i,l_i}(\eta,k_{i}) j_{l_{i}}(k_{i}r) \notag \\
&\times 
f^{(ij)}(k_{i})\bigg] \mathcal G^{m_{1}m_{2}m_{3}}_{l_{1}l_{2}l_{3}}\left(\mathcal G^{m_{1}m_{2}m_{3}}_{l_{1}l_{2}l_{3}}\right)^{*}\notag \\
&
=(4\pi)\sum_{j}\prod_{i=1}^{3}\left[\int \frac{dk_{i} k_{i}^{2}}{2\pi^{2}}\sum_{l_{i}} A_{i,l_i}(\eta,k_{i}) 
f^{(ij)}(k_{i})\right]\notag \\
&\times X_{l_{1}l_{2}l_{3}}J_{l_{1}l_{2}l_{3}}(k_{1},k_{2},k_{3}),\label{triple:result}
\end{align}
where we have used Eqs.~(\ref{Fourier:int}), (\ref{separable}) and partial wave expansion
\begin{align}
	e^{i\mathbf k\cdot \mathbf x}=4\pi\sum_{LM}i^{L}j_{L}(kr)Y_{LM}(\hat k)Y^{*}_{LM}(\hat x),
\end{align}
$j_{L}$ being the spherical Bessel functions.
Note that the Gaunt integral is also introduced as
\begin{align}
\mathcal G^{m_{1}m_{2}m_{3}}_{l_{1}l_{2}l_{3}}\equiv\int d\mathbf n \prod_{i=1}^{3}Y_{l_{i}m_{i}}(\mathbf n).	
\end{align} 
We derived the last line by defining 
\begin{align}
J_{l_{1}l_{2}l_{3}}(k_{1},k_{2},k_{3})
&\equiv \int^{\infty}_{0} dr r^{2}j_{l_{1}}(k_{1}r)j_{l_{2}}(k_{2}r)j_{l_{3}}(k_{3}r),	\\
X_{l_{1}l_{2}l_{3}}
&\equiv
4\pi\sum_{m_{1}m_{2}m_{2}}{}
\mathcal G^{m_{1}m_{2}m_{3}}_{l_{1}l_{2}l_{3}}\left(\mathcal G^{m_{1}m_{2}m_{3}}_{l_{1}l_{2}l_{3}}\right)^{*}.
\end{align} 
Then we use a shortcut notation to simply express the triple product as
\begin{align}
	&\hat {\mathcal F}\left[\prod_{i=1}^3A_{i,l_i}\right]\notag \\
	&\equiv (4\pi)\sum_{j}\prod_{i=1}^{3}\left[\int dk_{i} k_{i}^{2}(2\pi^{2})^{-1} 
f^{(ij)}(k_{i})A_{i,l_i}\right]
\notag \\
	&
	\times  J_{l_{1}l_{2}l_{3}}(k_{1},k_{2},k_{3}). \label{res:triple}
\end{align}
$X_{l_{1}l_{2}l_{3}}$ can be concretely evaluated as follows up to the quadruple moment:
\begin{align}
&\{X_{l_{1}l_{2}0},X_{l_{1}l_{2}1},X_{l_{1}l_{2}2}\}
\notag \\
&=\left\{
\left(\begin{array}{ccc}1 & 0 & 0 \\0 & 3 & 0 \\0 & 0 & 5\end{array}\right),
\left(\begin{array}{ccc}0 & 3 & 0 \\3 & 0 & 6 \\0 & 6 & 0\end{array}\right),
\left(\begin{array}{ccc}0 & 0 & 5 \\0 & 6 & 0 \\5 & 0 & \frac{50}{7}\end{array}\right)\right\}.\label{Xl123}
\end{align}
Note that we drop higher order multipole moments through out this paper for the same reason for Eq.~(\ref{yevol}).
Using Eqs.~(\ref{triple:result}) and (\ref{Xl123}), let us compute the ensemble average of the isotropic component of Eq.~(\ref{kdis:evo}).
The third-order collision term $\mathcal D$ was derived in Ref.~\cite{Ota:2016esq}, but angular dependence in Fourier space was not treated correctly.
Then we newly find the following expression:
\begin{align}
&\dot\tau^{-1}
 \llangle \mathcal D\rrangle =\hat{\mathcal F}\left[
3\Theta_{0}\Theta_{1g}V_{1}+6\Theta_{1}\Theta_{2}V_{1}\right]
+
\llangle V y\rrangle.
\end{align}
Eqs.~(\ref{coltemplin}), (\ref{triple:result}) and (\ref{Xl123}) yield
\begin{align}
&\frac12\dot\tau^{-1}
\llangle \Theta^{2}\mathcal A\rrangle = 
\hat{\mathcal F}\Big[\frac92\Theta_{0}\Theta_{2}^{2}
+\frac{45}{14}\Theta_{2}^{3}
\notag \\
&
-3\Theta_{0}\Theta_{1g}V_{1}
+\frac{87}{10}\Theta_{1}^{2}\Theta_{2}
-6\Theta_{1}\Theta_{2}V_{1}\Big].
\end{align}
Employing Eqs.~(\ref{triple:result}), (\ref{Xl123}) and (\ref{B2}), we also find
\begin{align}
&\dot\tau^{-1}
\left \llangle \Theta\mathcal B\right \rrangle=
\hat{\mathcal F}\Big[-3\Theta_{0}V_{1}^{2}
-6\Theta_{2}V_{1}^{2}+\frac{9}{2}\Theta _{0}\Theta^{2}_{2}\notag \\
&
+\frac{45}{14}\Theta^{3}_{2}
+3\Theta_{0}\Theta^{2}_{1}
+\frac{87}{10}\Theta^{2}_{1}\Theta_{2}
\Big]
-\frac{1}{ 4\pi }\langle \Theta_{00} y_{00}\rangle
\notag \\
& +\llangle \Theta y\rrangle
-\frac{1}{10\cdot 4\pi}\sum_{m=-2}^{2}\langle \Theta^{*}_{2m} y_{2m}\rangle.
\end{align}
Finally, the remaining part $y\mathcal A$ is
\begin{align}
&\dot\tau^{-1}
\left \llangle y\mathcal A\right \rrangle=
-\frac{1}{ 4\pi }\langle \Theta_{00} y_{00}\rangle +\llangle \Theta y\rrangle 
\notag \\
& 
-\frac{1}{10\cdot 4\pi}\sum_{m=-2}^{2}\langle \Theta^*_{2m} y_{2m}\rangle
-\llangle yV\rrangle.
\end{align}
Combining these expressions, we find
\begin{align}
&\frac{d\llangle\bar \kappa \rrangle}{d\eta} =-2\llangle y \mathcal A\rrangle
.\label{kmono}
\end{align}
Thus, the triple products of $\Theta$ are canceled, and only the mode coupling between $y$ and $\mathcal A$ contributes to $\kappa$.
The absence of triple products of the temperature multipoles in Eq.~(\ref{kmono}) implies that $y$ is necessary to produce $\kappa$.
In other words, $\kappa$ appears as a result of multiple scattering.
Now it is manifest that a possible source of the RHS of Eq.~(\ref{kmono}) is primordial non-Gaussianity.

\begin{figure}
\flushleft
\includegraphics[width=\linewidth]{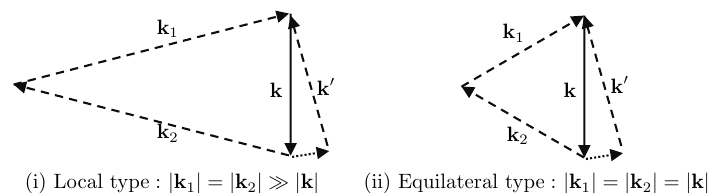}\caption{
Hierarchy of the scales.
The dashed arrow corresponds to the Fourier momenta in the convolutions of $\Theta^2_{2}$ or $\Theta^2_{1g}$.
The solid and dotted arrows are those of the $y$ and $\kappa$, respectively.
For the left squeezed shape, the superhorizon $y$ is produced from $\mathbf k_{1(2)}$ modes in the earlier stage.
Then, $y$ enters the horizon and produces $\kappa$ of $\mathbf k+\mathbf k'$ modes.
For the equilateral shape, $y$ and $\kappa$ are produced from $\mathbf k_1$, $\mathbf k_2$ and $\mathbf k'$ modes simultaneously.
In this case, our assumption behind Eq.~(\ref{localy}) is no more available and we need to account thoroughly for the nonlinear evolution of $y$.
In any case, we consider $|\mathbf k+\mathbf k'|\to 0$ limit when we calculate the ensemble average of $\kappa$.}
\label{shape}
\end{figure}

\subsection{Numerical estimation of $\kappa$}
Full evaluation of Eq.~(\ref{kmono}) requires full nonlinear evolution of the second-order $y$~\cite{Pitrou:2009bc}, but this is beyond the scope of this paper.
Instead, we roughly estimate $\kappa$ in a more simplified way.
First, we assume the local type configuration for primordial non-Gaussianity,
\begin{align}
	B_{\zeta}(k_{1},k_{2},k_{3})=\frac{6}{5}f^{\rm loc.}_{\rm NL}\left(P_{\zeta}(k_{1})P_{\zeta}(k_{2})+{\rm 2~perms.} \right).
\end{align} 
Then, we assume that $y$ has been already generated on superhorizon in the earlier epoch and that the nonlinear evolution in sub-horizon is negligible; we linearly interpolate free streaming of $y$ by employing the evolution equation without the source.
This approximation can be justified as long as we consider i) the local form non-Gaussianity enhanced in the squeezed limit and ii) the late period $z \sim 10^3$ because $y$ generation starts from $z\sim 5\times 10^4$.
Here, we write the initial superhorizon $y$ as $\zeta_{\mathbf k}^{y}$, which is obtained by integrating Eq.~(\ref{realspydif}) up to $z\sim 10^{3}$.
Note that $\langle \zeta^{y}\rangle=\llangle y\rrangle$ is satisfied in real space.
Then, transfer functions of $y$ can be introduced as we do in Eq.~(\ref{Fourier:int}):
\begin{align}
y(\eta,\mathbf k,\mathbf n)
\approx (4\pi)\sum_{lm}(-i)^{l}Y^{*}_{lm}(\mathbf n) Y_{lm}(\hat k)y_{l}(\eta,k)\zeta_{\mathbf k}^{y},\label{localy}
\end{align}
where the initial condition on superhorizon is given as $y_l=\delta_{l0}~(k\eta \ll 1)$. 
The statistics of $\zeta^{y}$ in Fourier space is calculated as
\begin{align}
\langle \zeta^{y}_{\mathbf k}\rangle &
\approx (2\pi)^{3}\delta(\mathbf k)\llangle y\rrangle ,\\
\langle \zeta^{y}_{\mathbf k}\zeta_{\mathbf k'}\rangle &
\approx (2\pi)^{3}\delta(\mathbf k+\mathbf k')\llangle y\rrangle\frac{12}{5}f^{\rm loc.}_{\rm NL}P_{\zeta}(k),\label{stat:yzeta}
\end{align}
where this approximation is valid if $|\mathbf k|$ is much smaller than $|\mathbf k_{1,2}|$, which are the Fourier momenta in the convolutions of $\Theta^2_2$ and $\Theta^2_{1g}$.
The relation between these momenta is depicted in the left panel of Fig.~\ref{shape}. 
The transfer function of $y$ is obtained by solving the following hierarchy equation without the source:
\begin{align}
&\dot y_l +\frac{k(l+1)}{2l+1} y_{l+1}-\frac{kl}{2l+1} y_{l-1}\notag \\
&=\dot\tau\left(1-\delta_{l0}-\frac{1}{10}\delta_{2l}\right)y_l.\label{y:heierarchy}
\end{align}

\begin{figure}
\flushleft
\includegraphics[width=\linewidth]{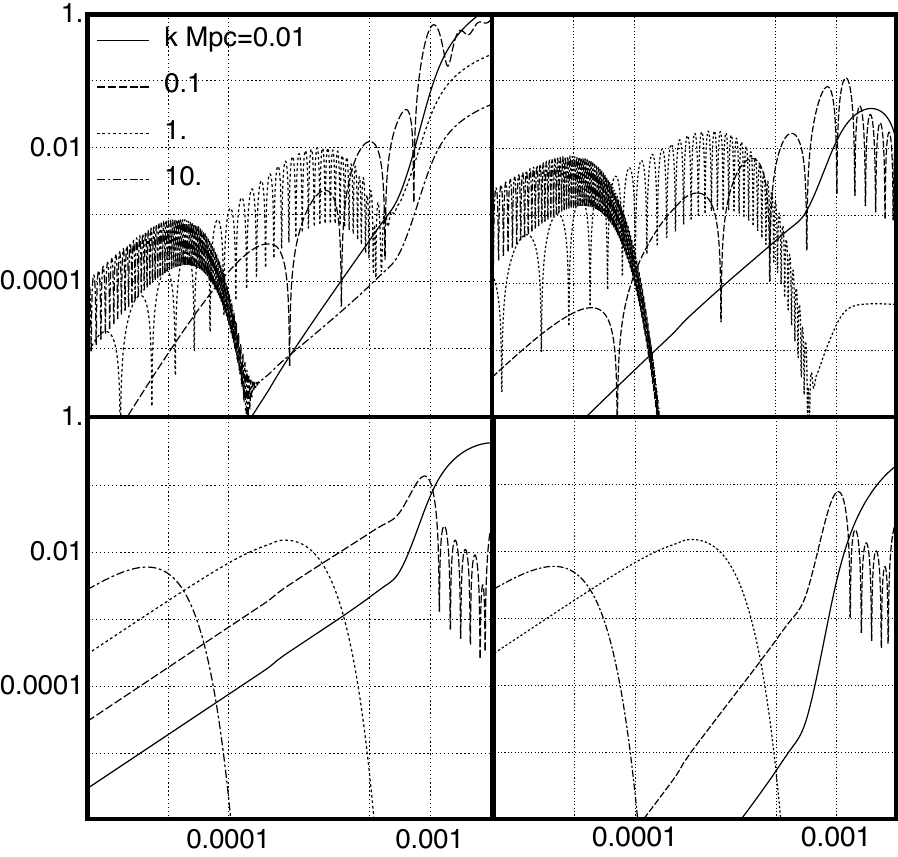}\caption{Transfer functions for $\Theta_{1}-V_{1}$~(top left), $\Theta_{2}$~(top right), $y_{1}$~(bottom left) and $y_{2}$~(bottom right).
The horizontal axis is the redshift.
In contrast to the temperature multipoles, $y$ multipoles do not oscillate in the earlier epoch because $y$ does not contribute to the velocity of photons because of the frequency dependence of $\mathcal Y$.
}
\label{transfer}
\end{figure}

Up to $l=2$, Eqs.~(\ref{coltemplin}), (\ref{kmono}) and (\ref{localy}) 
yield 
\begin{align}
&\llangle\bar\kappa\rrangle \approx -f^{\rm loc.}_{\rm NL}\llangle y\rrangle \int \frac{dk}{k}\frac{k^{3} P_{\zeta}(k)}{2\pi^{3}}
\notag \\
&\times\frac{24 }{5} \int^{\eta_{0}}_{\eta_{i}}d\eta \dot \tau \left[\frac{9}{2}\Theta_{2}y_{2}+3\Theta_{1g}y_{1}  \right].\label{kmonotransf}
\end{align}
Thus, gauge invariant variables like shear and heat conduction produce $\bar \kappa$.
Fig.~\ref{transfer} shows time evolution of $\Theta_{2}$, $\Theta_{1g}$, $y_{1}$ and $y_{2}$ calculated by modifying the cosmic linear anisotropy solving system (\texttt{CLASS})~\cite{Blas:2011rf}.
$y$ is erased in the earlier epoch when the universe is in kinetic equilibrium since they are converted into $\mu$.
We similarly account for such a thermalization effect for $\kappa$ by inserting 
\begin{align}
	J_{y}=\frac{1}{1+\left[\frac{1+z}{6\times 10^{4}}\right]^{2.58}},
\end{align} 
into Eq.~(\ref{kmonotransf}), assuming the same discussions for $y$~\cite{Chluba:2013vsa}.
Then, we numerically integrate Eq.~(\ref{kmonotransf}).
Fig.~\ref{kwindowplot} shows the estimation of the second line of Eq.~(\ref{kmonotransf}).
Though the Fourier space window function for $y$ picks modes on $k\, {\rm Mpc}\lesssim 10^{2}$ up~\cite{Chluba:2012we}, the contribution to $\kappa$
only comes from the modes on $k$Mpc$<$0.5.
This is because the phase discrepancy between $\Theta$ and $y$ cancels most of the energy injection.
Still, integration between $0.01<k\, {\rm Mpc}<0.5$ results in non zero value
\begin{align}
\llangle \bar\kappa\rrangle \approx -1.4\times 10^{-18}f^{\rm loc.}_{\rm NL}
\left(\frac{\llangle y\rrangle}{4\times 10^{-9}}\right),
\end{align}
where we set $k^{3}P_{\zeta}/2\pi^{2}=A_{\zeta}(k/k_{0})^{n_{s}-1}$ with $A_{\zeta}10^{9}=2.2$, $k_{0}$Mpc=0.05 and $n_{s}=0.96$.
Thus, $\kappa$ is directly related to primordial non-Gaussianity.

\begin{figure}
\flushleft
\includegraphics[width=\linewidth]{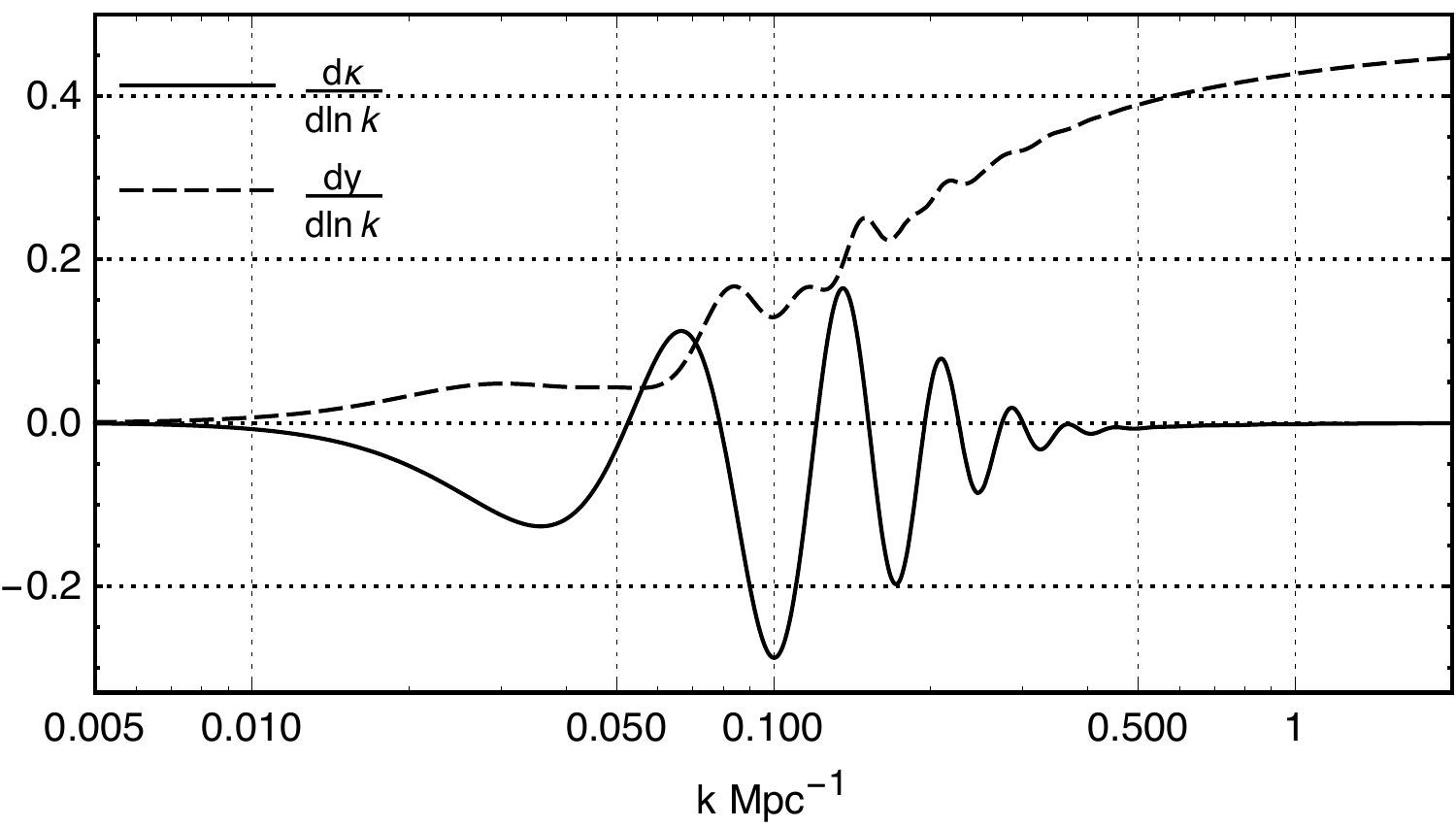}\caption{The Fourier space window functions for the spectral distortions in units of $f^{\rm loc.}_{\rm NL}\llangle y\rrangle k^{3} P_{\zeta}(k)/2\pi^{3}$~(solid line) and $ k^{3} P_{\zeta}(k)/2\pi^{3}$~(dashed line).
}
\label{kwindowplot}
\end{figure}

\section{Discussions}
\label{sec:discussion}
Even though the overall signal from primordial local non-Gaussianity is expected to be tiny, such a signal can, in principle, be distinguished from other types of CMB spectral distortions due to the specific frequency dependence of $\mathcal K$.
Note that we easily translate observational upper bounds on $y$ into those for $\kappa$, using $\int dp~p^{3}\mathcal K=4\int dp~p^{3}\mathcal Y$.
For example, the upper bound given by a primordial inflation explorer like experiment~\cite{Kogut:2011xw} is roughly estimated as $f^{\rm loc.}_{\rm NL}<\mathcal O(10^{8})$.
Notice that this bound is for squeezed non-Gaussianity whose short modes are on $1.<k{\rm Mpc}<100.$ since $y$ is produced on those scales, which cannot be constrained by the CMB temperature bispectra.
Though the signal might be extremely small, there are various aspects related to this new signal for the future investigations.
For example, the right panel of Fig.~\ref{shape} suggests that $\kappa$ is also sensitive to equilateral type non-Gaussianity, though this would require us a more exact estimation since the approximation behind Eq.~(\ref{localy}) is not valid.
Anisotropy in $\kappa$ would also be a new window for the primordial higher-order correlations. It is conceivable that the new cubic spectral distortion in Eq.~(\ref{kmono}) could also receive non-primordial contributions (e.g., weakly non-linear effects and projection effects, similarly to~\cite{Cabass:2018jgj}).  Finally, we expect astrophysical applications in the similar direction of multiple scattering for the Sunyaev-Zel'dovich effect~\cite{Sazonov:1998ae,Itoh:2001kj}.
Our result suggests that there exists a new type of spectral distortion if incoming photon distribution deviates from the ideal Planck distribution.
Therefore it is foreseeable that this process might also take place within clusters of galaxies.

\begin{acknowledgments}
We would like to thank Giovanni Cabass, Jens Chluba, Michele Liguori, Andrea Ravenni, Masahide Yamaguchi and Matias Zaldarriaga for useful discussions.
We would like to thank Jens Chluba, Enrico Pajer, Andrea Ravenni for careful reading of our manuscript.
We thank CERN for hosting the TH Institute ``Probing Fundamental with CMB Spectral Distortions'', where part of this work has been carried out. A.O. thanks the Physics and Astronomy Dept.~of Padova for their hospitality during the development of this work. 
A.O. is supported by JSPS Overseas Research Fellowships. N.B. acknowledges partial financial support by ASI Grant No. 2016-24-H.0.
\end{acknowledgments}


\end{document}